# First Light Sources at the End of the Dark Ages: Direct Observations of Population III Stars, Proto-Galaxies, and Supernovae During the Reionization Epoch

A White Paper Submitted to the Astro2010 Decadal Survey Committee

**Authors:**

Jeff Cooke (University of California, Irvine)
Asantha Cooray (University of California, Irvine)
Ranga-Ram Chary (California Institute of Technology)
Volker Bromm (University of Texas, Austin)
Renyue Cen (Princeton)
Richard Ellis (California Institute of Technology)
Elizabeth Fernandez (University of Colorado at Boulder)
Steven Furlanetto (University of California, Los Angeles)
Avi Loeb (Harvard University)
Anna Marie Moore (California Institute of Technology)
Leonidas Moustakas (Jet Propulsion Laboratory)
Peng Oh (University of California, Santa Barbara)
Brian O'Shea (Michigan State University)
Evan Scannapieco (Arizona State University)
Britton Smith (University of Colorado at Boulder)
Michele Trenti (University of Colorado at Boulder)
Aparna Venkatesan (University of San Francisco)
Daniel Whalen (Los Alamos National Laboratory)
Naoki Yoshida (Institute for the Physics and Mathematics of the Universe, Japan)

**Contact Information:**

**Dr. Jeff Cooke**
Department of Physics & Astronomy
University of California, Irvine
Frederick Reines Hall
Irvine, CA 92797-4575
(949)824-9257
cooke@uci.edu

**Executive Summary:**

The cosmic dark ages are the mysterious epoch during which the pristine gas began to condense and ultimately form the first stars. Although these beginnings have long been a topic of theoretical interest, technology has only recently allowed the beginnings of observational insight into this epoch. Many questions surround the formation of stars in metal-free gas and the history of the build-up of metals in the intergalactic medium:

- What were the properties of the first stellar and galactic sources to form in pristine (metal-free) gas?
- When did the epoch of Population III (metal-free) star formation take place and how long did it last?
- Was the stellar initial mass function dramatically different for the first stars and galaxies?

These questions are all active areas of theoretical research. However, new observational constraints via the direct detection of Population III star formation are vital to making progress in answering the broader questions surrounding how galaxies formed and how the cosmological properties of the universe have affected the objects it contains.

**Introduction:**

A wealth of data from cosmological surveys now suggests that the process of reionization was initiated within a few 100 Myr of the Big Bang and was completed by $z\sim6$. The sources of UV photons responsible for reionization, however, have yet to be identified observationally. Given the pristine nature of the intergalactic medium, consisting almost entirely of hydrogen and helium, the very first stars were metal-free (Population III). Their formation and evolution hold key clues that can establish how the Universe was subsequently enriched with metals, and how and when massive galaxies containing Population II stars formed. There is an apparent chemical enrichment "floor" of [Fe/H]$\sim$-4 in the solar neighborhood indicating a rapid, prompt enrichment of the IGM at high redshift. Constraining the initial mass function (IMF) of primordial stars is the key to understanding this rapid chemical enrichment, as well as primitive galaxy assembly, the reionization of the early cosmos, and possibly the origin of the supermassive black holes found at the centers of most massive galaxies today. Major inroads into the current observational desert ($7 < z < 20$), the epoch of reionization, can only be made by the next generation of ground-based and space-based instruments. In this *White Paper*, we outline a program involving multiple approaches to detecting early, metal-free stars, their supernovae, and the first galaxies.

In support of this observational program is the well-established theoretical community. Numerical simulations suggest that these primordial stars may exceed 100 solar masses (e.g., Bromm, Coppi, & Larson 1999, 2001; Abel, Bryan, & Norman 2000; Yoshida et al. 2006; Turk et al. 2008). The photon spectra of these stars produce hydrogen- and helium-ionizing photons with great efficiency (Bromm, Kudritzki, & Loeb 2001). Hydrodynamical simulations of the early universe complemented with radiative transfer suggest that metal-free stars form in small ($\sim 10^6$ solar mass) halos, or primordial proto galaxies (e.g., Abel et al. 2002; Bromm et al. 2002; O'Shea & Norman 2007).

The strongest direct spectroscopic signature of these hot stars in the observed-frame optical and near-infrared is likely intense HeII (1640 Å) emission, which would indicate a hard ionizing radiation field typical of a top-heavy stellar initial mass function and characteristic of Population III star formation. This feature can also be excited by AGN activity or by accretion on to mini-black holes, which could confuse the interpretation. Thus, direct detection of the *very* first stars and proto-galaxies is a difficult problem. Such stars may be easiest to find from their early supernova explosions, possibly visible in their rest-frame ultraviolet continua with the *James Webb Space Telescope (*JWST*)* or planned ground-based infrared surveys. Spectroscopy of the resultant supernovae with ground-based giant segmented mirror (infrared) telescopes (e.g., GSMT) should reveal metal abundance patterns typical of pair-creation supernovae providing unambiguous confirmation of the Population III nature of the progenitor stars.

Because chemical enrichment is a localized process, the first stars certainly did not enrich the entire universe all at once. Thus, even at "modest" redshifts ($4 < z < 7$), halos containing primordial gas might still exist and harbor Population III star formation. These epochs can also be probed to reveal the signatures of metal-free gas.

**Theoretical Models of Population III Star Formation:**

One of the most active pursuits in theoretical astrophysics is the creation of hydrodynamic simulations and cosmological arguments aimed at modeling the "first stars" (e.g., Bromm, Coppi, & Larson 1999, 2001; Abel, Bryan, & Norman 2000; Furlanetto & Loeb 2005; Yoshida et al. 2006; Turk et al. 2008; and many others). Although there are a lot of uncertainties involved, recent simulations have begun to suggest a solution in which the bulk of these early stars are massive (> 100 solar masses) and therefore extremely hot (Bromm, Kudritzki, & Loeb 2001). Figure 1 shows the predictions of the expected spectral energy distribution of these stars in the rest frame, illustrating the production of a large number of ionizing photons per unit star formation (Shaerer & Pelló 2002) and the resulting emission-line features. This efficient ionizing photon production and the accompanying hard radiation field provide the clearest expected observational signatures of this process. Detecting these signatures will provide much-needed constraints for models that describe pristine star formation in the universe.

As these early stars explode as highly energetic supernovae and enrich the surrounding intergalactic medium, the dramatic early epoch of star formation will ultimately transition into the Population II star formation that is more common today. Theoretical studies of this transition indicate that it is extended in time, and that some localized Population III star formation continues even to redshifts $z < 7$ (e.g., Furlanetto & Loeb 2005; Choudhury & Ferrara 2007; see Figure 1).

**Discovery of Supernovae from Early Stars:**

The earliest massive stars are predicted to end their lives as pair-instability supernovae (PISNe; e.g., Heger & Woosley 2002). Because these explosions involve huge amounts of energy ($10^{51}$-$10^{53}$ ergs), they may prove distinct and readily detectable with GSMTs.

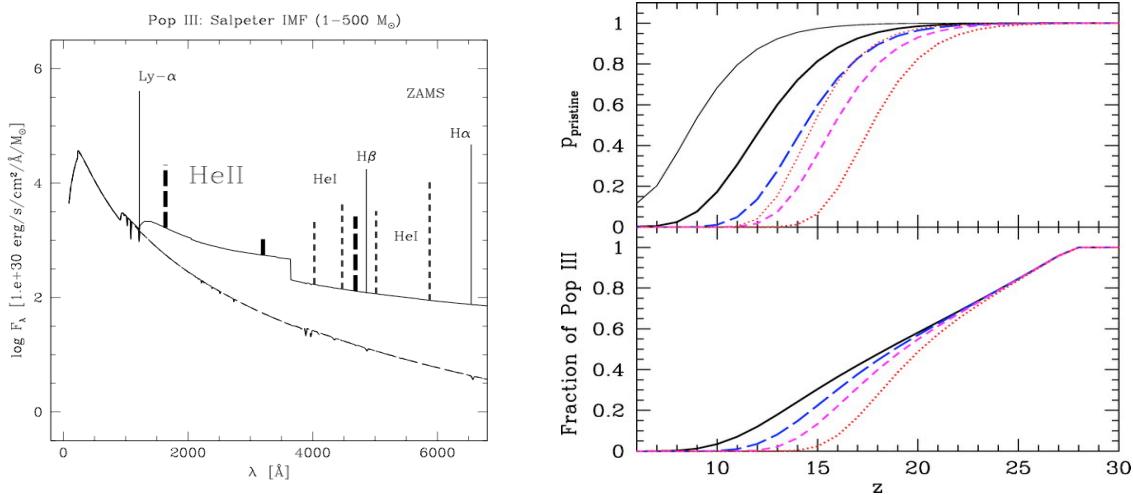

**Figure 1.** Models of population III star formation. (*Left*) Model spectral energy distributions of massive, metal-free (Population III) stars from Schaerer & Pelló (2002) showing the wealth of signatures of the "hard" radiation field produced by these massive and hot stars.
(*Right*) A cosmological model of the contribution of Population III star formation from Furlanetto & Loeb (2005). The top panel shows the fraction of halos that form with metal-free gas for various assumptions, and the bottom panels indicate the fraction of star formation that is Population III, all as a function of redshift.

Although it is currently hard to predict their exact properties, these early supernova explosions may have characteristics of Type IIn supernovae (SNe IIn), which are the deaths of massive (> 60 solar mass) stars (Gal-Yam et al. 2007; Smith et al. 2007) in the more local universe. These SNe are very luminous events, with rest-frame B-band luminosities as high as $M_B \sim -19$ (Richardson et al. 2002). Moreover, SNe IIn are the most luminous SN type in the rest-frame UV, rendering them easier to detect at high redshift than any other SN type. Observationally, SNe IIn at z~6 will be photometrically detected in deep, wide-field surveys that monitor high-redshift galaxies over multiple epochs for flux variations consistent with SN events (Cooke 2008). Application of this method to existing surveys has produced spectroscopically confirmed detections to z = 2.4 (Cooke et al. 2009).

There is a growing body of evidence that at least a fraction of SNe IIn are energetic pair-instability SNe (PISNe). As a result, SNe IIn provide the only observable examples of this process and have the potential to yield enormous insight into the behavior of high-mass (> 140 solar mass) Population III stars. Perhaps the best candidate for a PISN in the local universe is SN IIn 2006gy (Ofek et al. 2007; Smith et al. 2007, 2008). With a maximum brightness of $M_B$=-22.2, and estimated $M_{1500Å}$= -21, this event would have a magnitude of ~27 (AB) at z~6, rendering it detectable in deep imaging surveys. For example, the planned Hyper-SuprimeCam Deep layer will detect ~5-10 z~6 SN 2006gy events and a 10-year imaging program with the LSST will detect ~40,000 z > 2 SNe IIn and ~500 SN 2006gy events, with ~50 at z~6.

Once photometrically detected, core-collapse and PISNe that likely arise from more massive stars may be discriminated by their light-curve rise times and emission-line signatures. The late-time luminosities of core-collapse SNe IIn are largely driven by circumstellar interaction. They exhibit typical rest-frame UV and optical outburst rise times of ~15 days, both locally (e.g., Fransson et al. 2005) and at high redshift (Cooke et al. 2009), yet show more rapid decay rates in the rest-frame UV (~0.15 mag day$^{-1}$) as compared to optical (~0.03 mag day$^{-1}$). In contrast, the late-time optical luminosities of the highly energetic PISNe are largely powered by intense (> 10 solar mass) $^{56}$Ni decay. This process results in slow rise times of ~70 days in the rest-frame optical and decay rates of 0.01-0.2 mag day$^{-1}$. As a result, imaging surveys with a minimum of ~2 year baselines will be required fully acquire and discriminate between PISN and core-collapse SN light curves at z~6.

SNe IIn are defined by the presence of extremely luminous, long-lived emission lines that are dominated by Lyman α, MgII, and Hα. The ejecta from core-collapse SNe IIn interact with dense, previously expelled circumstellar material and remain strong for ~2 years, rest-frame, before beginning a slow decay. The emission lines from PISNe, however, are expected to drive the late-time luminosity for a shorter period of time as a result of the interaction of the ejecta with less dense stellar material expelled by intense winds. Emission lines from z~6 core-collapse SNe IIn are expected to be above the spectroscopic threshold of GSMTs for 10-15 years after outburst (see Fig. 2). SNe IIn line measurements using GSMTs will not only confirm photometric SN classifications but also reveal vital information regarding SN energies and metal abundance patterns. We note, however, that Lyman α will only be detectable above z ~ 6-7 when these sources reside inside ionized bubbles within the intergalactic medium (see below).

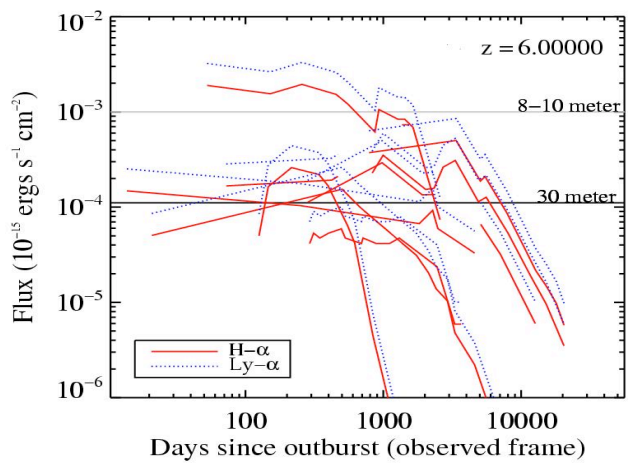

**Figure 2.** Emission-line strength evolution of local SNe IIn redshifted to z~6 from Cooke (2008). These SNe are thought to be the closest analogs to the pair-instability SNe expected from Population III star formation. At z~6 and beyond, the emission lines of SNe IIn may be detectable using 25-30-meter class GSMTs.

The combination of sensitive broad-band near-infrared imaging and spectroscopic follow-up with JWST, JDEM or ground-based GSMTs will provide the means to distinguish PISNe from core-collapse SNe by their light curve rise times, overall energy output, and relative emission-line characteristics. The longevity of the emission lines makes their detection and study conventional observations that require a baseline of 2-10 years. These are *not* target of opportunity observations. Undertaking searches for SNe IIn out to high redshift will reveal the evolving fraction of PISNe and thereby reveal the evolution of the fraction of baryons that are processed through Population III stars.

Constraining the evolution of the stellar IMF is arguably one of the most important topics in extragalactic astronomy. Currently, conclusions in favor of an evolving stellar IMF between 0 < z < 4 are drawn from the potential discrepancy between measurements of the stellar mass density and integral of the star-formation rate density (Hopkins & Beacom 2004; Davé 2008).

The latter quantity is inferred from ultraviolet and far-infrared emission, which are dominated by the most massive stars, while the stellar mass of galaxies is dominated by lower mass stars. The fact that the integral of the star-formation rate density appears to be a factor of 2 higher than the stellar mass density seems to suggest that the IMF in star-forming galaxies is top-heavy (but see Reddy et al. 2009 for a dissenting view). Measurements of the SN IIn rate as a function of redshift can settle this argument in a direct and unbiased way since the rates of such SNe would differ by a factor of 4 between the standard IMF scenario and the top-heavy IMF scenario. Wide field surveys in the optical/near-infrared bands on 4-m class telescopes will select obscured SNe, providing an unbiased measure of the SN IIn rate.

**Direct Detection of Population III "Dwarf" Galaxies**:

In Population III sources, direct detection of the HeII (1640 Å) emission line, expected with an equivalent width that exceeds ~10 Å, is within the reach of ground-based 25-30-meter telescopes. Figure 3 illustrates simple predictions for the He II counts of various star-forming scenarios described in Barton et al. (2004) and based on the calculations of Schaerer (2003) and Bromm et al. (2001). The "optimistic" scenario corresponds to pure Population III, metal-free, 300-1000 solar-mass stars. Both the "optimistic" and "plausible" scenarios can be detected with next-generation large ground-based telescopes. In reality, observed high-redshift sources will consist of contributions of Population II and Population III stars. With a combination of broadband imaging from JWST and GSMTs, plus deep spectroscopy from GSMTs, it is possible to measure the relative *fraction* of these populations, thereby providing a key constraint on cosmological models (e.g., Furlanetto & Loeb 2005; Choudhury & Ferrara 2007; Tornatore et al. 2007).

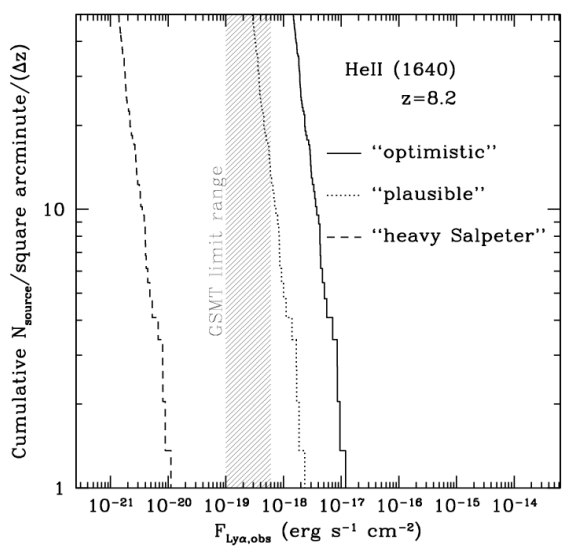

Figure 3. Simple model estimates of the HeII (1640 Å) "luminosity function" at $z \sim 8$. Following Barton et al. (2004), we plot the expected source counts in He II per square arcminute per redshift increment near $z \sim 8$ for fully Population III ("optimistic"), low-metallicity ("plausible"), and heavy Population II ("heavy Salpeter") scenarios. We also show the approximate sensitivity range of GSMTs. The presence of strong, detectable HeII emission is a clear indication of a drastic change in the stellar initial mass function and metal content of high-redshift galaxies.

At "modest" redshifts ($z < 11$) where Lyman α emission appears in the J-band, one of the best environments to discover Population III dwarf galaxies may be the chemically unevolved surroundings of a large galaxy or proto-cluster of galaxies. The central sources are likely to have ionized a local "bubble" through which strong Lyman α emission can escape from surrounding dwarf galaxies. Meanwhile, in the outskirts of the cluster, local enrichment may not yet have begun.  Thus, near-infrared observations of these dwarfs could reveal strong Lyman α and He II emission detectable by large ground-based telescopes, and possibly a rest-frame ultraviolet continuum observable from the ground and/or with the JWST.

By 2016, the JWST and/or ground-based surveys will regularly deliver deep infrared images that reveal the most massive star-forming galaxies and proto-clusters at these redshifts. Photometric redshifts, combined with spectroscopic follow-up with JWST or GSMT, will allow the identification the most luminous sources at specific redshifts in the range 7 < z < 11, such as z ~ 7.7, where Lyman α falls in an atmospheric window between night sky lines.  By using narrow band imaging (R ~ 300 – 1000) with modest-field (perhaps ≥ 15" to 2') near-infrared imagers on a GSMT, the environs of the massive systems that are of order the size of ionized "bubbles" (Mpc scales, or ~3.3´ at these redshifts) can be mapped to search for intense Lyman *a*. The actual length scale over which these sources are discovered will also reveal, or set a lower limit on, the size of the local bubble.  An exposure time of 4 hours will allow detections in the range of ~6 x $10^{-19}$ erg/s/cm$^2$ for 25-30-meter class telescopes.

When strong Lyman α emitters are found, both the Lyman α and the HeII lines can be observed with R>3000 spectroscopy using a GSMT. Observations can focus on the region in which HeII is expected (e.g., at 1.44 μm or H-band, for the z~7.7 window). With GSMTs, the likely sensitivity to unresolved emission at R > 3000 is ~1.5 x $10^{-19}$ erg/s/cm$^2$ for a half-night exposure.  If a HeII (1640 Å) feature is discovered that is strong relative to the UV continuum limits from JWST and the Lyman α feature (e.g., rest-frame EW (HeII) > 10-20 Å), the strength of the feature provides direct evidence that the source is not a "standard" Population II star formation region similar to what is observed in the local universe.  For example, even Wolf-Rayet stars in a standard stellar population cannot produce HeII equivalent widths in excess of a few Angstroms (Schaerer 2008).  Thus, the observations will provide compelling evidence for a very hard radiation field from pristine, Population III star formation. In addition, if strong HeII is present, a diffraction-limited IFU observation with a 25-30-meter telescope equipped with adaptive optics can determine whether the emission is spatially extended.  If it is resolvable and if, for example, a strong CIV feature is not also observed, the HeII emission likely does not arise from an AGN. By measuring the redshift evolution of the Lyman α luminosity function compared to the Lyman break galaxy luminosity function, we will have a unique window into the evolution of the neutral hydrogen fraction at the late stages of reionization.

These early, distant epochs of the universe are extremely elusive with present-day facilities because of a lack of sensitivity.  Current telescopes will likely discover larger populations of Lyman α sources.  However, with discovery limits in the vicinity of a few x $10^{-18}$ erg/s/cm$^2$, 8-10-meter class telescopes may not reveal Lyman α emission from dwarf galaxies at z ~ 7, unless the sources are gravitationally lensed.   As a result, this investigation is essentially impossible in the era before 25-to-30-meter telescopes.

**Indirect Detection of Population III: Efficient Production of Ionizing Photon**:

The WMAP measurement of the Thomson scattering optical depth, detection of GRBs at *z* > 6, deep cosmological surveys with Spitzer and Hubble, and ground-based studies of the Lyman α luminosity function at z > 5 are leading us close to a comprehensive picture of galaxy formation in the reionization epoch. Yet, there seems to be a fundamental discrepancy – the stellar mass density in galaxies at $z\sim6$ seems to indicate that the star-formation at $z > 6$ was insufficient to reionize the IGM to an extent which would reproduce the WMAP measurements (Chary 2008). The immediate implication of this result is that the average stellar IMF at $z > 6$ must be top-heavy with a functional form of $dn/dM\sim M^{-1.7}$, compared to a Salpeter slope of -2.3. This may either due to a transition from massive Population III stars to more normal Population II stars or due to different physical conditions in the temperature and gas density in the early Universe that favor the formation of massive stars. Are we already seeing the remnants of Population III star-formation in z~6 galaxies or was the epoch of zero metallicity star-formation so brief and rapid that a more traditional Salpeter IMF of stars was in place within 1 Gyr of the Big Bang? A combination of high redshift SN searches, Lyman α emitter surveys, deep spectroscopy, and high resolution HI maps of the Universe that measure the sizes of Stromgren spheres at high redshift will be required to address this question.